# Double superconducting dome and triple enhancement of $T_c$ in the kagome superconductor $CsV_3Sb_5$ under high pressure

K. Y. Chen[1,2=], N. N. Wang[1,2=], Q. W. Yin[3=], Z. J. Tu[3], C. S. Gong[3], J. P. Sun[1,2*], H. C. Lei[3*], Y. Uwatoko[4], and J.-G. Cheng[1,2*]

[1]*Beijing National Laboratory for Condensed Matter Physics and Institute of Physics, Chinese Academy of Sciences, Beijing 100190, China*

[2]*School of Physical Sciences, University of Chinese Academy of Sciences, Beijing 100190, China*

[3]*Department of Physics and Beijing Key Laboratory of Opto-electronic Functional Materials & Micro-nano Devices, Renmin University of China, Beijing 100872, China*

[4]*Institute for Solid State Physics, University of Tokyo, Kashiwa, Chiba 277-8581, Japan*

= These authors contributed equally to this work.

E-mails: jpsun@iphy.ac.cn, hlei@ruc.edu.cn, jgcheng@iphy.ac.cn

## Abstract

$CsV_3Sb_5$ is a newly discovered $Z_2$ topological kagome metal showing the coexistence of a charge density wave (CDW)-like order at $T^* = 94$ K and superconductivity (SC) at $T_c = 2.5$ K at ambient pressure. Here we study the interplay between CDW and SC in $CsV_3Sb_5$ via measurements of resistivity and magnetic susceptibility under hydrostatic pressures. We find that the CDW transition decreases with pressure and experience a subtle modification at $P_{c1} \approx 0.6$-$0.9$ GPa before it vanishes completely at $P_{c2} \approx 2$ GPa. Correspondingly, $T_c(P)$ displays an unusual M-shaped double dome character with two maxima around $P_{c1}$ and $P_{c2}$, respectively, leading to a tripled enhancement of $T_c$ to about 8 K at 2 GPa. The obtained temperature-pressure phase diagram resembles those of many unconventional superconductors, illustrating an intimated competition between CDW-like order and SC. The competition is found to be particularly strong for the intermediate pressure range $P_{c1} \leq P \leq P_{c2}$ as evidenced by the broad superconducting transition and reduced superconducting volume fraction. This work not only demonstrates the potential to raise the $T_c$ of the V-based kagome superconductors, but also offers more insights into the rich physics related to the electronic correlations in this novel family of topological kagome metals.

The newly discovered Kagome metals $AV_3Sb_5$ (A = K, Rb, Cs) have attracted considerable attention due to the observation of many intriguing phenomena, including superconductivity (SC), nontrivial band topology, and charge order [1-4]. At ambient conditions, these materials crystallize into a layered structure with hexagonal symmetry (space group *P6/mmm*), consisting of alkali-metal A layer and V-Sb slab stacked alternatively along the *c*-axis [1]. The most prominent feature of this structure is the presence of quasi-two-dimensional ideal kagome layers of V ions coordinated by Sb. These compounds are found to be metallic and even enter a superconducting ground state below the transition temperatures of $T_c$ = 0.93 K, 0.92 K, and 2.5 K for A= K, Rb, and Cs [2-4], respectively. Moreover, recent measurements of thermal conductivity on $CsV_3Sb_5$ single crystal at ultra-low temperature evidenced a finite residual linear term, pointing to an unconventional nodal SC [5]. Interestingly, proximity-induced spin-triplet SC and edge supercurrent were observed in the $Nb/K_{1-x}V_3Sb_5$ devices [6]. In addition, in the normal state they all exhibit a clear anomaly at $T^* \sim$ 78-104 K in the electrical transport and magnetic properties due to the formation of charge order (CDW-like) as revealed by the X-ray diffraction and scanning tunning microscopy measurements [2-4,7]. The charge order in $KV_3Sb_5$ has been found to display a chiral anisotropy [7], which can lead to giant anomalous Hall effect in the absence of magnetic order or local moments [8,9]. It has also been argued as a strong precursor of unconventional SC [7]. Moreover, angle-resolved photoemission spectroscopy measurements and density-functional-theory calculations have characterized their normal state as a $Z_2$ topological metal with multiple Dirac nodal points near the Fermi level [1,4]. The observations of pronounced Shubnikov-de Haas quantum oscillations and small Fermi surfaces with low effective mass in $RbV_3Sb_5$ single crystals are consistent with the existence of multiple highly dispersive Dirac band near the Fermi level [3]. In addition, similar to staggered arrangement of hexagonal graphene [10], theoretical calculations on the kagome Hubbard model with different electron fillings have shown the stabilization of many exotic phases, such as the spinless fermions [11-13], Mott transition [11,14,15], charge density waves [11,16,17], chiral spin-density-wave state [11], exotic superconducting state [11,12] and topological point defects [17]. Therefore, these V-based kagome metals $AV_3Sb_5$ have been regarded as a novel platform to study the interplay between SC, nontrivial band topology, and electronic correlations [1-9].

At present, the topologically related phenomena and SC have been actively studied in these $AV_3Sb_5$ compounds, but the possible rich physics related to the electronic correlation, especially the relationship between the intertwined electronic orders, has been barely revealed. In this regard, it is highly interesting to unveil the correlation between the CDW-like order and SC commonly observed in these $AV_3Sb_5$ materials. Here we have chosen to study $CsV_3Sb_5$ single crystal with the highest $T_c$ = 2.5 K among this series of compounds by applying the high-pressure approach, which has been widely used in disentangling the competing electronic orders of the strongly correlated metals [18].

Through detailed measurements of resistivity, DC and AC magnetic susceptibility on $CsV_3Sb_5$ single crystals under hydrostatic pressure, we successfully uncover an M-shaped double superconducting dome associated with the modification of the CDW-like order and its complete suppression under pressure, respectively. Interestingly, the $T_c$ of $CsV_3Sb_5$ is triply enhanced to about 8 K at 2 GPa, indicating that the $T_c$ of these V-based kagome superconductors can be raised considerably. By revealing the coexistence and competition of SC with the CDW-like order, the temperature-pressure phase diagram of $CsV_3Sb_5$ resembles those of many unconventional superconductors, and thus offer more insights into the rich correlation-related physics pertinent to this novel family of kagome metals.

Single crystals of $CsV_3Sb_5$ were grown from Cs ingot (purity 99.9%), V powder (purity 99.9%) and Sb grains (purity 99.999%) using the self-flux method, similar to the growth of $RbV_3Sb_5$ [3]. Temperature dependences of resistivity for two $CsV_3Sb_5$ samples (#1, #2) and ac magnetic susceptibility for sample #3 were measured simultaneously by using a self-clamped piston cylinder cell (PCC) under various hydrostatic pressures up to 2.2 GPa [19]. Daphne 7373 was used as the pressure transmitting medium (PTM) in PCC. The pressure values in PCC were determined from the superconducting transition of Pb according to the equation: $P$ (GPa) = ($T_0$ - $T_c$)/0.365, where $T_0$ = 7.20 K is the $T_c$ of Pb at ambient pressure. We also measured the temperature-dependent resistivity on sample #4 up to 6.6 GPa with a palm-type cubic anvil cell (CAC) [20]. Glycerol was employed as the liquid PTM for CAC. DC magnetization measurements on sample #5 under various pressures up to 1.06 GPa were also performed in a miniature BeCu PCC fitted into the commercial magnetic property measurement system (MPMS-3, Quantum Design). Details about the crystal growth procedures and high-pressure measurements can be found in the Supplementary Materials (SM).

Figures 1(a) and 1(b) show the temperature dependence of resistivity $\rho(T)$ and its derivative $d\rho/dT$ of sample #1 under various pressures up to 2.2 GPa measured with a PCC. These data illustrate clearly how the CDW-like order evolve with pressure. At 0 GPa, the CDW transition at $T^* \approx 94$ K is manifested as a kink-like anomaly in $\rho(T)$ and a corresponding peak in $d\rho/dT$, respectively. With increasing pressure, $T^*$ moves to lower temperature progressively and reaches about 15 K at 1.86 GPa, above which it cannot be discerned any more in resistivity, implying a complete suppression of CDW-like order by pressure. It is noteworthy that the kink feature in $\rho(T)$ at $T^*$ become much diminished at 0.6-0.9 GPa, and changes at $P \geq 1.2$ GPa to a hump-like anomaly, Fig. 1(a), i.e., the resistivity experiences an enhancement rather than a rapid decrease upon cooling through $T^*$. This is a typical character for the CDW transition due to the gap opening over part of the Fermi surfaces [21-23]. Accordingly, the anomaly in $d\rho/dT$ changes from a peak for $P < 0.6$ GPa through an intermediate crossover region 0.6-0.9 GPa to a dip for $P \geq 1.2$ GPa as shown in Fig. 1(b). These observations are confirmed in sample #2 measured in the sample PCC, as seen in Fig. S1 of SM. These results indicate that the CDW-like order at ambient pressure undergoes a subtle change and might evolve into a distinct CDW state at $P > 0.6$-$0.9$ GPa. Such a change at the normal state has a profound impact on the superconducting transition as shown below.

Figure 2(a) displays an enlarged view of the $\rho(T)$ data in Fig. 1(a) at $T \leq 10$ K, highlighting a complex, non-monotonic variation with pressure of the superconducting transition. As marked by arrows in Fig. 2(a), the onset of superconducting transition, $T_c^{onset}$, is determined from the cross point of two straight lines above and below the transition, while the $T_c^{zero}$ is defined as the zero-resistance temperature. The $\rho(T)$ at 0 GPa shows a relatively sharp superconducting transition with $T_c^{onset} \approx 3.5$ K and $T_c^{zero} \approx 2.8$ K, consistent with the previous result [4,5]. With increasing pressure, the normal-state resistivity is enhanced, and the $T_c^{onset}$ and $T_c^{zero}$ are raised up quickly to ~ 5 K and ~ 4.3 K at 0.37 GPa, and to ~ 6.8 K and ~ 5 K at 0.61 GPa, respectively. At 0.61 GPa, there exists a small step before reaching zero resistivity, indicating the presence of two superconducting phases. This small resistivity-step feature seems to evolve into a stronger shoulder at 0.9 GPa, resulting in a significantly broad superconducting transition width $\Delta T_c \approx 4.2$ K with the $T_c^{onset}$ ~7.2 K and the $T_c^{zero}$ ~ 3 K. When applying pressure to 1.2 GPa, the normal-state resistivity is reduced gradually, and the two-step feature of the superconducting transition is weakened by reducing the $T_c^{onset}$ to ~ 5.7 K but leaving the $T_c^{zero}$ at ~ 2.8 K, which is close to that at 0.9 GPa. But the transition width $\Delta T_c \approx 3$ K is still large. Interestingly, $T_c$ is shifted again to higher temperatures upon further increasing pressure; the $T_c^{onset}$ and $T_c^{zero}$ at 1.86 GPa reaches ~ 8 K and ~ 6 K, respectively. In this pressure range 1.2-1.86 GPa, the superconducting transition width $\Delta T_c \approx 2$-$3$ K is still relatively large, featured by either a broad onset or a long-tail when approaching zero. It should be noted that such a broad superconducting transition is not caused by the pressure inhomogeneity but is an intrinsic phenomenon, originating from the microscopic phase coexistence and competition as discussed below.

When the CDW-like order just vanishes at about 2 GPa, the superconducting transition becomes very sharp with a $\Delta T_c \approx 0.3$ K. As seen in Fig. 2(a), the values of $T_c^{onset}$ and $T_c^{zero}$ are 7.96 and 7.63 K for 2.03 GPa, and 7.68 and 7.43 K for 2.19 GPa, respectively. It seems that $T_c$ will decrease again at higher pressures. To track the evolution of $T_c(P)$ in a larger pressure range, we measured $\rho(T)$ of sample #4 up to 6.6 GPa with a CAC. The $\rho(T)$ data at low temperatures are shown in Fig. 2(b), and those in the whole temperature range are given in Fig. S2. The results in CAC are highly consistent with those in PCC. As can be seen, the $\rho(T)$ of 1.9 GPa in CAC resembles that of 1.86 GPa in PCC, showing a relatively sharp onset but a long tail with $T_c^{onset} \approx$ 7.8 K and $T_c^{zero} \approx 6.5$ K. When increasing pressure to 2.5 GPa and above, the superconducting transition becomes very sharp with $\Delta T_c \approx 0.3$ K and shifts to lower temperatures monotonically. The $T_c^{onset}$ and $T_c^{zero}$ are reduced to 2.82 and 2.58 K at 6.6 GPa.

Since $T_c(P)$ exhibits a complex variation with pressure and the $\Delta T_c$ is quite wide in the pressure range 0.6-1.9 GPa where the high-temperature CDW order coexists, it is essential to further characterize the superconducting transition via detailed magnetic measurements. To this end, we measured the dc magnetization $M(T)$ of sample #5 up to 1.06 GPa in MPMS and the ac susceptibility $\chi'(T)$ of sample #3 up to 2.2 GPa with mutual induction method in PCC. Figure 2(c) shows the $M(T)$ data collected upon warming up under an external magnetic field of $H$ = 20 Oe after zero-field cooled from room temperature. As can be seen, the diamagnetic signal in $M(T)$ appears at $T_c^M$ = 2.5 K for 0 GPa, in agreement with the resistivity data, and it moves quickly to ~ 6.3 K at 0.54 GPa. In this pressure range, the bulk nature of SC is also confirmed by the large diamagnetic response of $M(T)$. However, for $P$ = 0.74 and 0.88 GPa, the diamagnetic response in $M(T)$ appears at high temperature of ~ 6-7 K, but the transition is very broad and its magnitude is dramatically lowered. These observations suggest that the superconducting volume fraction is substantially reduced, consistent with the $\rho(T)$ data in the similar pressure range as shown in Fig. 2(a). At 1.06 GPa, a large diamagnetic signal emerges again below 3 K, signaling the resurgence of a new bulk superconducting phase.

The $\chi'(T)$ results in Fig. 2(d) show one-to-one correspondence to the $\rho(T)$ data shown in Fig. 2(a), including the two-step feature of $\chi'(T)$ at 0.61 and 0.9 GPa (inset of Fig. 2(d)), the reduction of $T_c$ from 0.6 to 1.2 GPa followed by a resurgence of $T_c$ up to 1.86 GPa with a relatively broad transition, and a sharp superconducting transition at $P \geq 2$ GPa. It is noteworthy that the superconducting volume fraction is relatively small, ~30-50%, in the pressure range 0.6 -1.86 GPa when the CDW order coexists, while a nearly 100% bulk SC is finally realized at 2 GPa when the CDW order vanishes completely. These results thus provide direct evidences for the microscopic coexistence and competition between CDW and SC [24].

Based on the above comprehensive high-pressure characterizations, we can construct the $T$-$P$ phase diagram of CsV$_3$Sb$_5$, Fig. 3(a, b), which depicts explicitly the evolution and correlations of $T^*$ and $T_c$ as a function of pressure. As can be seen, $T^*(P)$ decreases monotonically with pressure and vanishes completely around $P_{c2} \approx 2$ GPa, while $T_c(P)$ displays an M-shaped double-dome character with two maxima around $P_{c1} \approx 0.8\pm0.2$ GPa and $P_{c2}$, respectively. As shown in Figs. 3(b) and 3(c), the superconducting transition width $\Delta T_c$ in the pressure range $P_{c1} \leq P \leq P_{c2}$ is significantly larger that at the lower and higher pressures. The highest $T_c \approx 8$ K is achieved around $P_{c2}$ and it is three times higher than that at ambient pressure. This observation immediately calls attention to further raise the $T_c$ of these V-based kagome superconductors.

*Discussions*. The major finding of the present work is the observation of an M-shaped double superconducting dome that has an intimated interplay with the high-temperature CDW-like order. As shown in Fig. 1(a), the $\rho(T)$ anomaly around $T^*$ displays a subtle change from a kink-like rapid reduction at $P < 0.6$ GPa to a hump-like weak upturn at $P > 0.9$ GPa through an intermediate crossover region 0.6 - 0.9 GPa. Accordingly, the anomaly in $d\rho/dT$ around $T^*$ changes from a peak to a dip. To illustrate such a change, the symbols of $T^*$ in Fig. 3(a) are color-coded in terms of the sign of $\Delta(d\rho/dT)$ at $T^*$ as defined in Fig. 1(b), which shows a clear crossover around $P_{c1}$. This means that, although the suppression of CDW-like order by pressure leads to an initial enhancement of $T_c$, the modification of the CDW state around $P_{c1}$ shows a stronger competition with the superconducting pairing, leading to the first extremum of $T_c$ shown in Fig. 3(b).

As mentioned above, the weak upturn of $\rho(T)$ upon cooling through $T^*$ in the pressure range $P_{c1} \leq P \leq P_{c2}$ is a typical character for the formation of CDW-like state that opens a gap over part of the Fermi surfaces. Although the density wave instability in this class of kagome metals is not unexpected, this pressure-induced CDW order should be distinct from the ambient-pressure one given the different responses in $\rho(T)$. At first, it seems to fulfill a better nesting condition across the in-plane two-dimensional Fermi surface, which can be finely tuned by compression [25]. As a result, the resistivity exhibits an enhancement across $T^*$. Secondly, this new CDW order has a stronger tendency to compete for the electronic state responsible for SC [24,26], featured by a microscopic phase separation, leading to very broad superconducting transitions and substantially reduced superconducting volume fraction as observed in Figs. 2, and 3(b, c). The continuous suppression of the CDW order by pressure results in the resurgence of SC at higher pressures, leading to a second extremum of $T_c$ around $P_{c2}$ where the CDW order just vanishes.

However, the observed maximal $T_c$ in the vicinity of $P_{c2}$ followed by a subsequent monotonic reduction of $T_c$ at higher pressure do not conform the typical behaviors of conventional two-dimensional CDW superconductors, such as transition-metal dichalcogenides $Ta(Se,S)_2$, in which the $T_c(P)$ shows a plateau or broad spectrum even after the CDW is suppressed by doping or pressure [27-29]. Instead, the $T-P$ phase diagram of $CsV_3Sb_5$ shown in Fig. 3(a, b) resembles those of many unconventional superconducting systems, such as heavy-fermion [30], cuprates [31], iron-based superconductors [18,32-34] and $Lu(Pt_{1-x}Pd_x)_2In$ [35], which are characterized by the presence of quantum criticality.

To examine such a possibility, we probe the evolution of the electronic states by evaluating the upper critical field $\mu_0H_{c2}$ of the superconducting state. Figure S3 shows all the $\rho(T)$ data under various magnetic fields at different pressures for sample #1 in PCC and sample #4 in CAC. As can be seen, the superconducting transition is shifted gradually to lower temperatures with increasing magnetic fields as expected.

However, the critical field required to eliminate the superconducting transition exhibits a strong dependence as a function of pressure. In order to quantify this evolution, here we determined $T_c$ as the middle-point temperature of superconducting transition and plotted $\mu_0H_{c2}$ versus $T_c$ in Fig. 4. Then, we can estimate the zero-temperature upper critical field $\mu_0H_{c2}(0)$ by fitting the $\mu_0H_{c2}(T)$ with the empirical Ginzburg–Landau (GL) equation, viz. $\mu_0H_{c2}(T) = \mu_0H_{c2}(0)(1 − t^2)/(1 + t^2)$, where $\mu_0H_{c2}(0)$ is zero-temperature upper critical field and $t$ is the reduced temperature $T/T_c$. The fitting curves are shown as the broken lines in Fig. 4 and the extracted $\mu_0H_{c2}(0)$ values are plotted in Fig. 3(d) as a function of pressure. Interestingly, the $\mu_0H_{c2}(0)$ also displays two pronounced peaks around $P_{c1}$ and $P_{c2}$, respectively. The corresponding $\mu_0H_{c2}(0)$ values are larger than 3 T, which is about one order of magnitude higher than that at ambient pressure. Similar double-peak features are also observed in the pressure dependence of the initial slope $\mu_0H_{c2}(T)$, i.e., $-dH_{c2}/dT|_{Tc}$, which is proportional to the effective mass of charge carriers [36]. As shown in Fig. S4, the divergence behaviors of $-dH_{c2}/dT|_{Tc}$ around $P_{c1}$ and $P_{c2}$, especially the latter, signal the dramatic enhancement of effective mass, which has been regarded as a hallmark of quantum criticality [37]. The presence of quantum criticality around $P_{c2}$ is conceivable due to complete suppression of the CDW order, in line with many unconventional superconductors [38-43]. However, whether there is a buried quantum critical point around $P_{c1}$ deserves further investigations. To unveil the nature of the subtle change of the CDW-like order around $P_{c1}$ should be key to understand these peculiar behaviors.

## Conclusion

In summary, we performed a comprehensive high-pressure study on the electrical transport and magnetic properties of the $CsV_3Sb_5$ single crystal, which is a newly discovered $Z_2$ topological kagome metal showing the coexistence of a CDW-like order at $T^* = 94$ K and SC at $T_c = 2.5$ K at ambient pressure. Our results uncover a hitherto unknown pressure-induced modification of the CDW order around $P_{c1} \approx 0.6$-0.9 GPa before it is completely suppressed around $P_{c2} \approx 2$ GPa. Accordingly, $T_c(P)$ exhibits an unusual M-shaped double superconducting dome with two maxima occurring right at $P_{c1}$ and $P_{c2}$, respectively, thus revealing an intimated interplay between the CDW and SC. The competition between these two electronic orders is particularly strong for the intermediate pressure range $P_{c1} \leq P \leq P_{c2}$ as evidenced by the strong reduction of superconducting volume fraction and the broad transition width. The $T_c$ of $CsV_3Sb_5$ can be triply enhanced to ~ 8 K at a moderate pressure of 2 GPa, implying that the $T_c$ of these V-based kagome superconductors still has a room to go higher. In addition, the double-peak character has also been observed in the $\mu_0H_{c2}(0)$, and characteristics of quantum criticality around $P_{c1}$ and $P_{c2}$ has also been indicated. The determined T-P phase diagram with a quantum criticality around $P_{c2}$ resemble those of many unconventional superconductors, thus providing more ingredients related to the strong electron correlations into the rich physics of this novel family of topological kagome metals. Several open issues still need to be addressed in the future studies, such as the nature of the CDW-like order in the intermediate pressure range, and the plausible buried quantum critical point around $P_{c1}$.

## Acknowledgements

This work is supported by the National Natural Science Foundation of China (12025408, 11904391, 11921004, 11888101, 11834016, 11822412 and 11774423), the Beijing Natural Science Foundation (Z190008 and Z200005), the National Key R&D Program of China (2018YFA0305700, 2018YFE0202600 and 2016YFA0300504), the Strategic Priority Research Program and Key Research Program of Frontier Sciences of the Chinese Academy of Sciences (XDB25000000, XDB33000000 and QYZDB-SSW-SLH013), and the CAS Interdisciplinary Innovation Team.

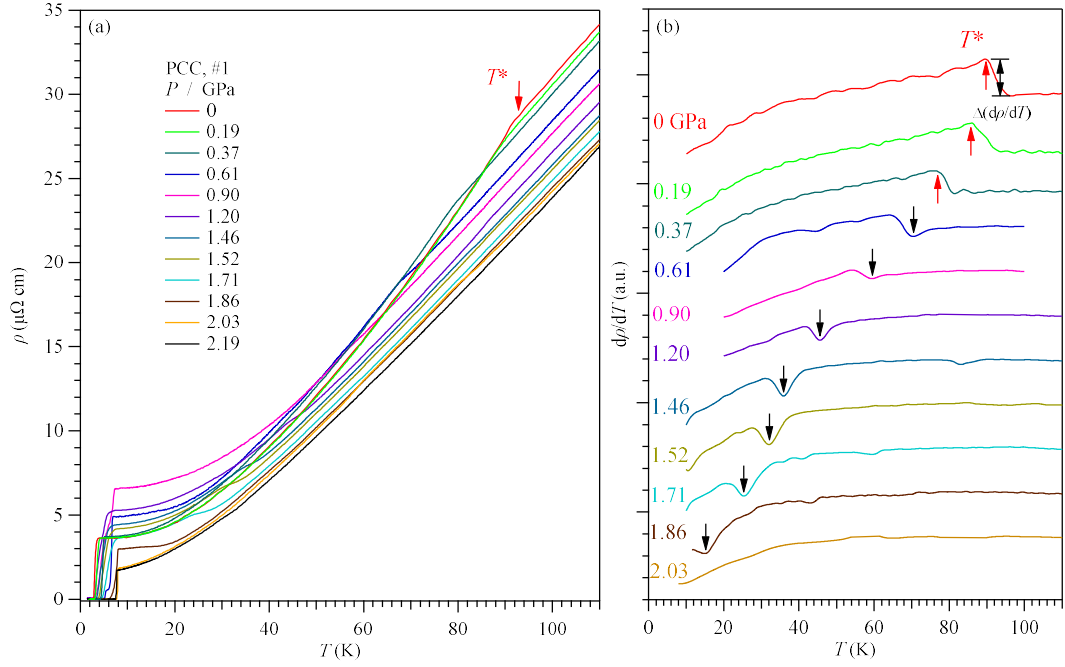

Figure 1. Variation with pressure of the CDW-like transition. Temperature dependences of (a) resistivity $\rho(T)$ and (b) its derivative $d\rho/dT$ for the $CsV_3Sb_5$ sample #1 measured in PCC under various pressures up to 2.19 GPa. The transition temperature of CDW-like order, $T^*$, are marked by the arrows in the figure. The curves in (b) have been shifted vertically for clarity.

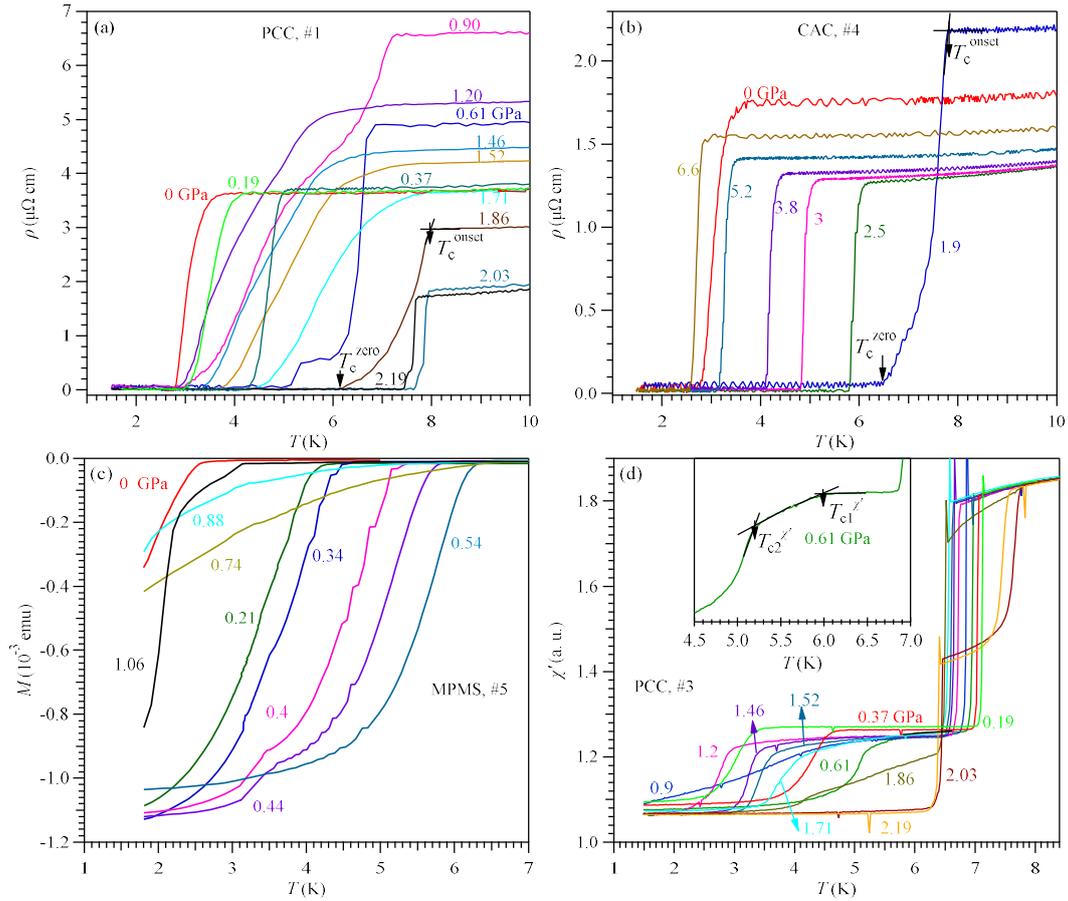

Figure 2. Variation with pressure of the superconducting transition in (a, b) resistivity and (c, d) magnetic susceptibility. The resistivity $\rho(T)$ data in (a) and (b) are measured for sample #1 up to 2.19 GPa with a PCC and for sample #4 up to 6.6 GPa with a CAC, respectively. The dc magnetization in (c) was recorded in MPMS on sample #5 with a miniature PCC, while the ac magnetic susceptibility in (d) was measured on sample #3 with the mutual induction method in PCC. Inset of (d) shows the curve at 0.61 GPa.

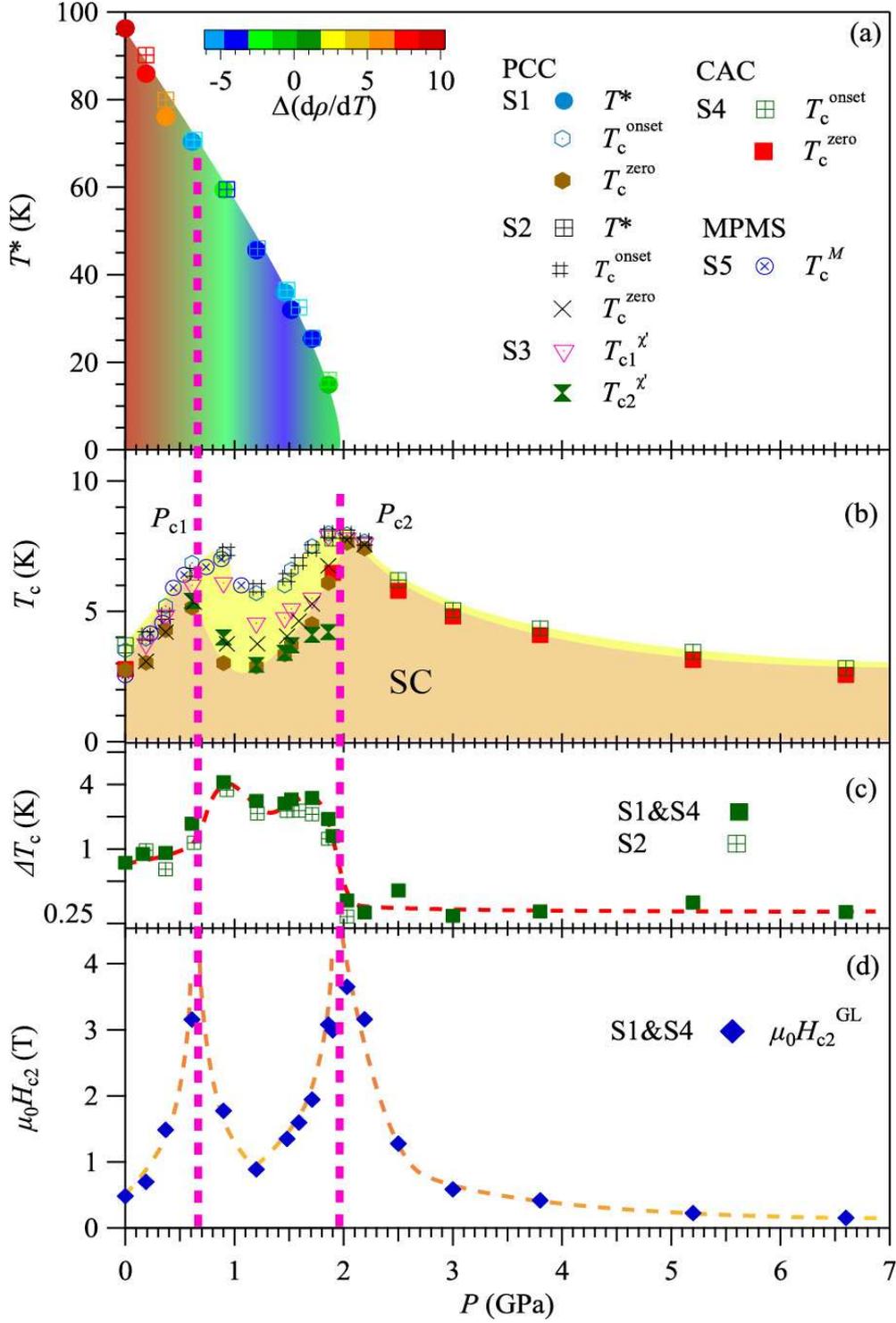

Figure 3. Temperature-pressure phase diagram of $CsV_3Sb_5$. Pressure dependences of (a) the CDW-like transition temperature $T^*$, (b) the superconducting transition temperatures $T_c^{onset}$, $T_c^{zero}$, $T_c^{\chi}$, and $T_c^M$ determined from the resistivity and magnetic measurements on several samples, (c) the superconducting transition width $\Delta T_c$, and (d) the zero-temperature upper critical field $\mu_0 H_{c2}(0)$ obtained from the empirical Ginzburg–Landau (GL) fitting.

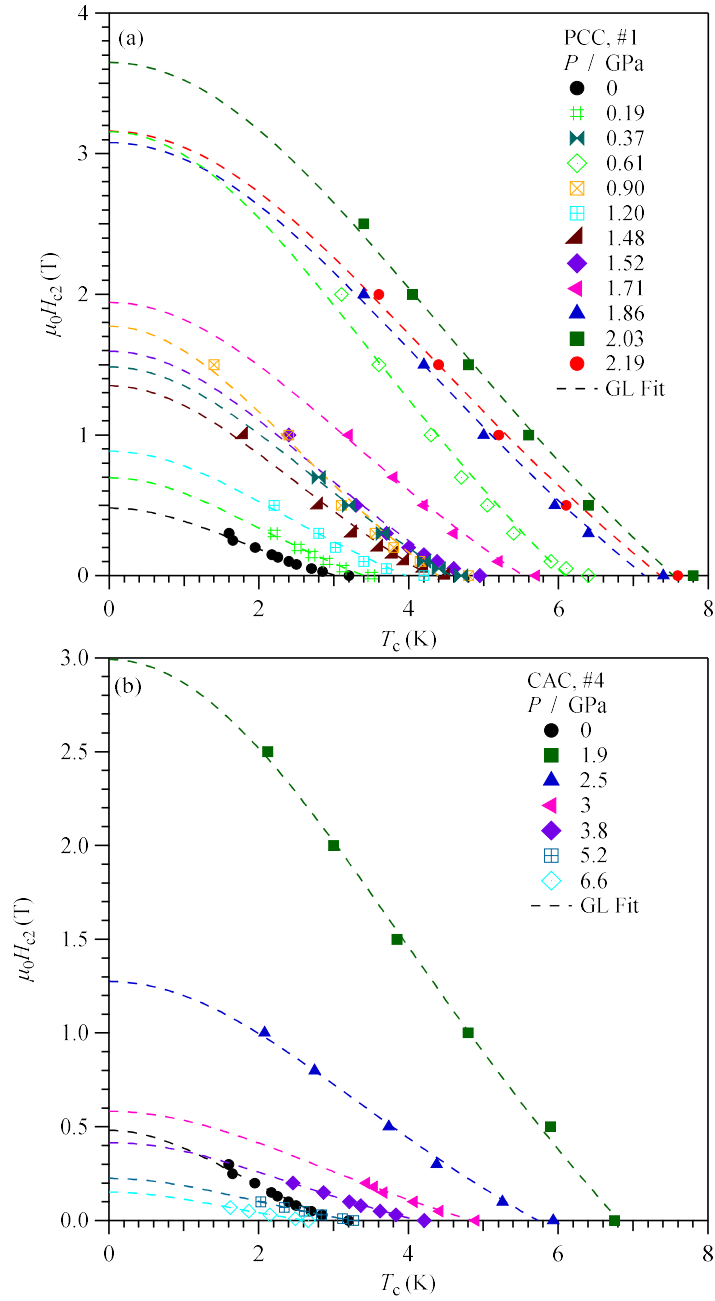

Figure 4. Temperature dependence of the upper critical field $\mu_0 H_{c2}$ at different pressures measured for (a) sample #1 with PCC and (b) sample #4 with CAC. The broken lines represent the Ginzburg–Landau (GL) fitting curves.

15